# Directional fidelity of nanoscale motors and particles is limited by the 2$^{nd}$ law of thermodynamics–via a universal equality

Zhisong Wang[1,2,3*], Ruizheng Hou[1,3], Artem Efremov[1]

[1]*Department of Physics,* [2]*NUS Graduate School for Integrative Sciences and Engineering,* [3]*Center for Computational Science and Engineering, National University of Singapore, Singapore 117542*

*Correspondence author (phywangz@nus.edu.sg)

ABSTRACT

Directional motion of nanoscale motors and driven particles in an isothermal environment costs a finite amount of energy despite zero work as decreed by the 2$^{nd}$ law, but quantifying this general limit remains difficult. Here we derive a universal equality linking directional fidelity of an arbitrary nanoscale object to the least possible energy driving it. The fidelity-energy equality depends on the environmental temperature alone; any lower energy would violate the 2$^{nd}$ law in a thought experiment. Real experimental proof for the equality comes from force-induced motion of biological nanomotors by three independent groups – for translational as well as rotational motion. Interestingly, the natural self-propelled motion of a biological nanomotor (F1-ATPase) known to have nearly 100% energy efficiency evidently pays the 2$^{nd}$-law decreed least energy cost for direction production.

# I. INTRODUCTION

Machines –big or small– must produce directional motion to do work. Nanomotors[1] or externally driven nanoscale particles typically execute directional motion in an isothermal environment as any temperature gradient in the immediate surrounding would be readily leveled by fast heat transfer over the small dimension. Directional motion in an isothermal environment is generally limited by the $2^{nd}$ law of thermodynamics as the law requires an energy cost to sustain the direction even if no work is done. The energy cost must be other than the environmental heat; otherwise a load might be attached to the moving object to draw work continually from heat of the single-temperature environment. This would turn the object into a perpetual machine of the second type, which violates the $2^{nd}$ law. Hence an energy price for pure direction exists and it must be above zero.

But what is the least price of pure direction for nanoscale motors and particles? Is it of any practical relevance to nanoscience and –technology? The answers are subject to how the 'pure direction' is quantified. Here we use a concept of fidelity to quantify motional direction of arbitrary nanoscale objects that are intrinsically stochastic. Rather surprisingly, the least energy price thereby formulated follows a universal equality, which is further verified by experiments reported by multiple independent groups on force-induced motion of biological nanomotors. Interestingly, the natural self-propelled motion of a biological nanomotor known to have ~ 100% energy efficiency evidently pays the $2^{nd}$ law-decreed least price for pure direction, and this channel of energy consumption accounts for a significant portion of the motor's total energy input.

The concept of directional fidelity was previously[2,3] used to study molecular motors, and a cycle analysis turned out to be useful in exposing connections[3] between a motor's directional fidelity and thermodynamic quantities like entropy productions. In this study, we generalize the fidelity concept to arbitrary nanoscale objects, and find a similar cycle analysis applicable to any sustained directional motion, be it induced[4,5,6] externally by an applied force/field or internally by a self-propelled motor. Further conceptual and quantitative developments for general isothermal motion lead to the equality linking to the 2$^{nd}$ law.

## II. METHODS

### A. General stochastic kinematics of nanoscale objects and directional fidelity

Directional fidelity is pertinent to any nanoscale objects because their motion is done by microscopic transitions that in principle are stochastic and reversible. In general, the stochastic kinematics of a directionally moving nanoscale object in an isothermal environment may be quantified against a frame of reference defined by the free energy versus position of the object along its direction of motion (see Fig. 1A for an example). The path of the object is characterized by the array of traversed free-energy minima where the object, prior to any energy input, is most probably found according to Boltzmann's law. The stochastic transitions activated by thermal fluctuations or/and an energy input generate not only forward displacements from one free-energy minimum to another but also backward displacements. Besides, null displacements occur in which a forward or backward displacement is initiated but not completed with the object instead returning to its original free-energy minimum. The all and only essence of direction of the moving object appears to be captured by the probability for net forward displacements. This probability is the directional fidelity, or alternatively called directionality[2,3], which is quantified as

$$D = (p_f - p_b)/(p_f + p_b + p_0). \qquad (1)$$

Here $p_f$, $p_b$ are the probabilities for forward and backward displacements between adjacent free-energy minima and $p_0$ is the probability for the associated null displacements. These displacements within adjacent free-energy minima are basic and can combine to account for any larger motion along the array of free-energy minima. The probabilities $p_f$, $p_b$ and $p_0$ can be obtained by averaging over many of the displacements of an object without knowing its driving mechanism. The directional fidelity thus defined is applicable to arbitrary nanoscale objects moving in an isothermal environment regardless of the mechanisms by which the direction is induced.

The direction is perfect with $D = 1$, and is lost entirely for $D = 0$ (i.e. pure Brownian motion in which the forward displacements cancel the backward ones). $D$ is similar to the Peclet number quantifying transport phenomena in fluid flows, but the former is capped by one and the latter may be above one. By counting null displacements too, $D$ also differs from the forward-to-backward stepping ratio $R = p_f/p_b$ and the ratio $(p_f - p_b)/(p_f + p_b)$ used in previous studies[7-10].

**B. Relevance of cycles to sustained directional motion**

For sake of a general discussion, the free-energy minima along the path of a directionally moving nanoscale object are called stations (marked by **A**), and the free-energy maxima between any two neighbouring stations are called bridges (marked by **B**). As shown in Fig. 1A, the object's motion along the array of stations may be traced back to four types of transitions between adjacent stations and bridges, namely a forward station-to-bridge transition (marked by rate $k_{AB}^+$) and the reverse backward transition ($k_{BA}^-$), plus a forward bridge-to-station transition ($k_{BA}^+$) and the reverse transition ($k_{AB}^-$). To sustain a directional motion, the transitions must form self-closed cycles that repetitively produce inter-station displacements by rounds of energy consumption. Four elemental cycles may be identified (Fig. 1A). $k_{AB}^+ \to k_{BA}^+$ forms a cycle (marked by cycle flux $J_c^+$) that produces forward displacements from one station to an adjacent station; $k_{AB}^- \to k_{BA}^-$ forms a cycle ($J_c^-$) producing reverse backward displacements. $k_{AB}^+ \to k_{BA}^-$ and $k_{AB}^- \to k_{BA}^+$ form cycles too ($J^0_{c\alpha}$ and $J^0_{c\beta}$), which but produce null displacements. A direction sustainable by a stable supply of energy is quantified by a steady-state directionality. As the probabilities for forward, backward and null displacements ($p_f$, $p_b$ and $p_0$) are respectively proportional to the cycle fluxes $J_c^+$, $J_c^-$, and $J^0_{c\alpha} + J^0_{c\beta}$, the steady-state directionality can be expressed in terms of steady-state fluxes of the four elemental cycles. Namely,

$$D = (J_c^+ - J_c^-)/(J_c^+ + J_c^- + J_{c\alpha}^0 + J_{c\beta}^0). \tag{2}$$

### C. Directional fidelity-entropy relation

For a periodic array of stations/bridges, an object's motion is sufficiently described by a transition diagram in terms of only four doorway states of a pair of adjacent station and bridge (Fig. 1B), although the station/bridge may accommodate more states. The doorway states are those by which the forward moving object enters and exits a station (marked A[in] and A[out]) or bridge (B[in] and B[out]). The transition diagram has four transition pathways linking four doorway states. As shown in Fig. 1C, the four cycle fluxes can be obtained by decomposing the transition fluxes accompanying the four transitions along the two station-bridge pathways (A[out]↔B[in] and B[out]↔A[in]): $J_{AB}^+ = J_c^+ + J^0_{c\alpha}$, $J_{BA}^- = J_c^- + J^0_{c\alpha}$, $J_{BA}^+ = J_c^+ + J^0_{c\beta}$, and $J_{AB}^- = J_c^- + J^0_{c\beta}$. Here $J_{AB}^+ = p_{A[out]} k_{AB}^+$ is the flux accompanying the A[out]→B[in] transition with $p_{A[out]}$ being the occupation probability for A[out] state; other three transition fluxes are similarly defined. This yields $D = (J_{BA}^+ - J_{AB}^-)/(J_{BA}^+ + J_{BA}^-)$. Applying steady-state condition $J_{AB}^+ + J_{AB}^- = J_{BA}^+ + J_{BA}^-$ further yields $D = (J_{AB}^+/J_{BA}^- - 1)(J_{BA}^+/J_{AB}^- - 1)/(J_{AB}^+J_{BA}^+/J_{BA}^-J_{AB}^- - 1)$. Note that the total entropy productions[11,12] of the object plus environment due to the net forward flux through the two station-bridge pathways are $\Delta S_{AB} = k_B \ln(J_{AB}^+/J_{BA}^-)$ and $\Delta S_{BA} = k_B \ln(J_{BA}^+/J_{AB}^-)$ ($k_B$ is Boltzmann constant). Replacing the flux ratios in *D* with the entropy productions yields an exclusive directionality-entropy relation,

$$D = \frac{(e^{\Delta S_{AB}/k_B} - 1)(e^{\Delta S_{BA}/k_B} - 1)}{e^{(\Delta S_{AB} + \Delta S_{BA})/k_B} - 1}. \tag{3}$$

**D. Directional fidelity-energy equality**

The amount of energy that is actually consumed to produce a net forward displacement is the sum of the environmental temperature (*T*) times the entropy productions accompanying the net forward fluxes along all the four transition pathways of the forward-displacement cycle ($J_c^+$).

$$\Delta G = T\Delta S_{AB} + T\Delta S_{BA} + T\Delta S_{AA} + T\Delta S_{BB}. \tag{4}$$

Here $\Delta S_{AA}$ and $\Delta S_{BB}$ are the entropy productions along the intra-station and intra-bridge transition pathways. Note that the temperature times an entropy production is heat. The energy consumption $\Delta G$ is all dissipated into heat irreversibly to sustain the resultant net forward fluxes along the forward-displacement cycle, i.e. a pure direction that is measured by the resultant *D*. Hence $\Delta G$ of eq. 4 is identified as the energy consumption sustaining a pure direction (*D*). Following the definitions of entropy productions and transition fluxes, $\Delta G$ is related to the forward-to-backward transition rate ratio along each pathway as $\Delta G = k_B T \ln[(k_{AB}^+/k_{BA}^-)(k_{BB}^+/k_{BB}^-)(k_{BA}^+/k_{AB}^-)(k_{AA}^+/k_{AA}^-)]$, which suggests $\Delta G$ as the total free-energy drop[12] actually coupled with the forward-displacement cycle. Hence for nansocale particles or motors in general, their energy consumption for *D* is identified as the irreversible heat generation during a full forward inter-station displacement.

The maximum direction out of a given amount of energy consumption ($\Delta G$) is obtained by maximizing $D$ in eq. 3 using the two entropy productions $\Delta S_{AB}$ and $\Delta S_{BA}$ as variables under the energy constraint of eq. 4. When $\Delta G$ is shared by only two entropy productions, i.e. $\Delta S_{AB} = \Delta S_{BA} = \Delta G/2T$, the optimization yields the maximum direction as $D_{max}(\Delta G) = \tanh(\Delta G/4k_B T)$. Namely,

$$D_{max}(\Delta G) = \frac{e^{\Delta G/2k_B T} - 1}{e^{\Delta G/2k_B T} + 1}. \tag{5}$$

Inversely, $\Delta G$ on the right-hand side of Eq. 5 may be interpreted as the minimum energy necessary to produce the direction on the left-hand side. Namely $\Delta G_{min}(D) = 4k_B T \times \text{artanh}(D)$, or in a more transparent form

$$\Delta G_{min}(D) = 2k_B T \ln\left(\frac{1+D}{1-D}\right). \tag{6}$$

The two limits $D_{max}(\Delta G)$ and $\Delta G_{min}(D)$ hold also for non-periodic stations/bridges. In this case, an ensemble of doorway states for the stations/bridges must be considered. The four-pathway transition diagram like Fig. 1B still applies; but each pathway contains many transitions, and its associated transition fluxes are each a sum over individual transitions and states. A cycle analysis for the case of non-periodic stations/bridges was previously done in ref.[3], yielding an upper bound for the overall $D$ that is the same as the fidelity-entropy relation eq. 3 except $\Delta S_{AB}$ and $\Delta S_{BA}$ being replaced by the highest entropy productions by individual pairs of reverse transitions in the two station-bridge pathways. Multiple transition pairs of differing entropy productions invariably yield $D$ below this upper bound (i.e. eq. 3) – thereby below $D_{max}(\Delta G)$ of eq. 5. Hence $D_{max}(\Delta G)$ is valid for non-periodic stations/bridges, and so is $\Delta G_{min}(D)$. Nevertheless, periodic stations/bridges are practically a convenient way to approach $D_{max}(\Delta G)$ and $\Delta G_{min}(D)$ because the $D$ upper bound of ref.[3] is readily recovered as eq. 3 with either station-bridge pathway reduced to a single transition pair (Fig. 1B).

## III. RESULTS AND DISCUSSIONS

### A. The fidelity-energy equality in force-induced directional motion

Validity of eqs. 5 and 6 may be examined for an arbitrary nanoscale particle being pulled by an external force in a periodic environmental potential (Fig. 2). In this case a single state may be assigned to a station or bridge, hence $\Delta S_{AA} = \Delta S_{BB} = 0$. Other entropy productions are $\Delta S_{AB} = \Delta G^+_{AB}/T + k_B\ln\gamma_{AB}$ and $\Delta S_{BA} = \Delta G^+_{BA}/T - k_B\ln\gamma_{AB}$. Here $\gamma_{AB} = p_A/p_B$ is a steady-state population ratio, $\Delta G^+_{AB} = k_BT\ln(k^+_{AB}/k^-_{BA})$ and $\Delta G^+_{BA} = k_BT\ln(k^+_{BA}/k^-_{AB})$ are free-energy gaps between adjacent bridges/stations. Following eq. 4, the energy consumption for the force-induced direction is the free-energy drop between two adjacent stations: $\Delta G = f{\times}d = \Delta G^+_{AB} + \Delta G^+_{BA} = T(\Delta S_{AB} + \Delta S_{BA})$ where $f$ is the pulling force and $d$ the inter-station separation. This $\Delta G$ produces the maximum $D$ of eq. 5 or becomes $\Delta G_{min}(D)$ of eq. 6 for the resultant $D$ when $\Delta S_{AB} = \Delta S_{BA} = \Delta G/2T$, which holds if $\Delta G^+_{AB} = \Delta G^+_{BA} = \Delta G/2$ and $\gamma_{AB} = 1$. A sufficient condition to meet the two requirements is an entirely flat environmental potential, which renders the particle free of any local binding along the path of motion, hence avoids dissipation associated with bond rupture[13]. Pulling such a particle with a force of a fixed direction is conceivably a most efficient scenario of direction induction if the force and resultant motion are collinear. A flat potential eliminates stations/bridges altogether and renders arbitrary the inter-station separation $d$; yet eqs. 5 and 6 remain valid. This further confirms generality of the two limits.

**B. The minimum energy for a directionality fidelity may be converted to work**

The minimum energy price $\Delta G_{\min}(D)$, though dissipated into entropy to produce $D$, may be converted to work by 100% when $D$ is nullified by a resisting load. Consider the above ideal case of a pulled particle in a flat potential. A load $F$ opposing the particle's directional motion reduces the pull to $f_{\text{eff}} = f - F$. The direction generated by the reduced pull, $D_{\text{eff}}$, is again given by eq. 5, except $\Delta G$ being replaced by $\Delta G_{\text{eff}} = f_{\text{eff}} \times d$. Since $\Delta G_{\min}(D) = f \times d$, $\Delta G_{\text{eff}} = \Delta G_{\min}(D) - F \times d$. Increasing load to $F_s = \Delta G_{\min}(D)/d$ reduces $\Delta G_{\text{eff}}$ to zero; and $D_{\text{eff}}$ becomes zero too by eq.2. Hence $F_s$ is the stall force at which the direction vanishes. Under $F = F_s$ the particle achieves the maximum work, which is exactly the minimum energy for the initial direction, namely $W_{\max} = F_s \times d = \Delta G_{\min}(D)$.

**C. The equality captures the absolute least energy price decreed by the 2$^{\text{nd}}$ law**

Is $\Delta G_{\min}$ of eq. 6 the least energy price for a direction $D$ in general? If this were not true, a smaller amount of energy, $\Delta G'$ ($< \Delta G_{\min}$) would produce the same level of $D$ in a sustainable way. Nullifying such a $D$ can produce work up to $\Delta G_{\min}$ by the above analysis. Thus the energy $\Delta G'$ would produce the amount of work $\Delta G_{\min}$. This would violate the energy conservation if no heat of the environment is converted to work. If a finite amount of heat, $Q = \Delta G_{\min} - \Delta G' > 0$, of the thermally equilibrated environment is converted to work so as to satisfy the energy conservation, the 2nd law of thermodynamics would be violated because a cyclic heat-to-work conversion would occur under a single temperature. Thus, retaining $D$ by an energy less than $\Delta G_{\min}(D)$ violates the 2$^{\text{nd}}$ law, and must be falsified for any possible mechanism of direction induction. This Gedanken experiment concludes that the energy price given by eq. 6 is the absolute least universally as long as the 2$^{nd}$ law forbids any cyclic heat-to-work conversion in a single-temperature environment.

As a least energy price decreed by the 2$^{nd}$ law, $\Delta G_{min}(D)$ in unit of environmental temperature times the Boltzmann constant ($k_BT$) is exclusively linked to directionality without any explicit dependence on induction mechanisms, geometry/energy of the stations/bridges, transition rates, speed and step size (i.e. inter-station distance). Such an exclusive energy-directionality relation tends to qualify $\Delta G_{min}(D)$ as the 2$^{nd}$ law-decreed least energy cost for pure direction and qualify $D$ as an appropriate quantification of the pure direction: the size independence of $\Delta G_{min}(D)$ separates it from pure work that depends explicitly on the step size; the speed independence separates $\Delta G_{min}(D)$ from pure dissipation that depends explicitly on speed. Nevertheless, $\Delta G_{min}(D)$ overlaps with both work and dissipation: it is entirely an entropy-generating dissipation to sustain $D$ when no work is done, but contributes to work when it is done against a $D$-reducing load. This mixed role further relates $\Delta G_{min}(D)$ to the channel of energy consumption for pure direction as direction for a motion is necessary to produce work against a load and heat against friction. $\Delta G_{min}(D)$ recovers the limit of directionless Brownian motion (i.e., $\Delta G_{min} = 0$ at $D = 0$), and of cost-free direction in absolute vacuum (i.e., $\Delta G_{min} \to 0$ at $T \to 0$). $\Delta G_{min}(D)$ becomes infinite for perfect direction ($D = 1$) due to stochastic nature of microscopic motion. This does not contradict the common observation of seemingly perfect direction for macroscopic objects: the associated energy consumption (e.g. inevitable dissipation by friction) becomes macroscopic too, typically on the magnitude of Avogadro number times $k_BT$ (i.e. $\Delta G \sim 6.02 \times 10^{23} k_BT$); yet a $\Delta G$ of mere 100 $k_BT$ already affords a virtually perfect $D$ of $1 - 3.8 \times 10^{-22}$!

**D. Experimental proof of the fidelity-energy equality**

An ideal testing ground for the equality is low-dissipation directional motion of submicroscopic objects induced by an external force. As previously discussed, such a motion has a chance to approach the least price by a collinear coupling between the force and the resultant motion. High-resolution data of force-induced collinear motion are reported[8-10] for biological nanomotors F1-ATPase (a rotational motor) and kinesin (a translational motor along a linear track), which both move by steps of a fixed size[14,15] with a low[6,9,13,14] dissipation associated with bond rupture inside the motors and at the motor-substrate interface. The stepwise pattern of motion allows directionality ($D$) and associated energy consumption ($\Delta G$) to be deduced from the applied force ($F$) and the measured steps. Both motors make forward steps along their natural direction; but a large opposing force $F$ induces a reverse motion if $F$ overcomes the maximal resisting force produced by either motor (namely the so-called stall force/torque $F_s$). The energy consumption per step is the force surplus ($F - F_s$) times the step size ($d$), namely $\Delta G = (F - F_s) \times d$.

The displacement probabilities $p_f$, $p_b$, $p_0$ for F1 and kinesin motors are not available from experiments, but their directionality can be deduced from the so-called forward-to-backward stepping ratio ($R$). In the single-motor experiments yielding $R$ for $F_1$[9] and kinesin[8,15,16], the detected forward and backward steps are not cycle-enabled full inter-station displacements (i.e., $J_c^+$, $p_f$ and $J_c^-$, $p_b$). This is made clear by a recent $F_1$ experiment[9]. Theoretical studies[3,17] suggest that the detected steps are actually individual displacements that can be associated with the station-bridge pathways. Hence the measured stepping ratio is not $R = J_c^+/J_c^-$, but $R = J_{AB}^+/J_{BA}^-$ or $J_{BA}^+/J_{AB}^-$ or $R = (J_{AB}^+ + J_{BA}^+)/(J_{BA}^- + J_{AB}^-)$ depending on which pathway produces major detectable displacements. Following the relations between transition fluxes and cycle fluxes, the three interpretations are $R = (J_{c+} + J_{c0,\beta})/(J_{c-} + J_{c0,\beta})$, $R = (J_{c+} + J_{c0,\alpha})/(J_{c-} + J_{c0,\alpha})$ or $R = (J_{c+} + J_{0/2})/(J_{c-} + J_{0/2})$ with $J_{0/2} = (J_{c0,\alpha} + J_{c0,\beta})/2$. Thus the detected forward and backward steps both contain futile steps, in consistence with the $F_1$ experiment[9]. These interpretations become equivalent and converge to the same $D$-$R$ relation as $D = (R-1)/(R+1)$ if the direction induction is close to the least price: $\Delta S_{AB} = \Delta S_{BA}$ leads necessarily to $J_{AB}^+/J_{BA}^- = J_{BA}^+/J_{AB}^-$ and thereby $J_{c0,\alpha} = J_{c0,\beta} = J_{0/2}$. Validity of the $D$-$R$ relation for F1 and kinesin is alternatively established by the fact that futile steps are rare [9,15] for both motors ($p_0 \ll p_f$). Ignoring futile steps ($p_0$) leads to the same $D$-$R$ relation but as an approximation: now $R \approx p_f/p_b$, $D = (p_f - p_b)/(p_f + p_b + p_0) \approx (p_f - p_b)/(p_f + p_b)$, hence $D \approx (R-1)/(R+1)$. This is a good approximation because the absolute error from ignoring futile steps is $\delta D \approx p_0 \times (p_f - p_b)/(p_f + p_b + p_0)^2$ that is small at either high or low $F$. A high $F$ produces a good direction ($p_0 \ll p_f$) hence a small error; a low $F$ produces a negligible direction ($p_f - p_b \rightarrow 0$) and thereby a small error too.

For both F1-ATPase and kinesin, the force-induced $D$ versus $\Delta G$ measured at various $F$ values matches the $\Delta G_{min}(D)$ equality despite the data fluctuations typical of the single-molecule experiments (Fig. 2B, C). The data were collected by three independent groups[8-10] from three different motor systems, yet all follow the trend of the equality. Notably, the agreement is found not only for natural kinesin but also for a mutant[10] whose stall force ($F_s$) is reduced from ~ 7 pN to 4 pN, 3.2 pN and 2.4 pN depending on operation conditions. The latter value ($F_s = 2.4$ pN) is already close to the viscous drag[13] (~ 1 pN). The overall agreement of the diverse experiments with the same equality is unlikely accidental – it is rather an experimental proof for the equality and its general applicability to microscopic motion – translational as well as rotational.

**E. The least energy price in natural self-propelled motion of biological nanomotors**

Interesting enough, F1-ATPase, in its natural self-propelled forward motion, also pays the least energy price for direction production. For this motor-induced motion, the energy consumption per step is the chemical energy from the fuel consumption ($\Delta\mu$) minus the work done against a small opposing torque ($F < F_s$), namely $\Delta G = \Delta\mu - Fd$. The data[9] of $\Delta G$ versus $D$ for F1-ATPase's natural forward motion against various $F$ match the $\Delta G_{min}(D)$ equality (Fig. 3). The data indicate $R \approx 273$ for a low torque $F \approx 5$ pN nm/rad, and an extrapolation yields $R \approx 1497$ for zero-torque operation (see fig. 6B of ref.[9]). These results suggest an fidelity and its associated price as $D \approx 99.27\%$, $\Delta G_{min}(D) \approx 11.22$ $k_B T$ for $F \approx 5$ pN nm/rad and $D \approx 99.88\%$, $\Delta G_{min}(D) \approx 14.84$ $k_B T$ for $F = 0$ pN. The two values of $\Delta G_{min}(D)$ are ~ 67-74% and ~ 89-98% of the fuel energy[9] ($\Delta\mu \approx 15.89 \pm 0.75$ $k_B T$), suggesting the energy price for pure direction as a major channel of energy consumption for F1-ATPase's operation at low load. And this motor's energy for zero-load $D$ virtually accounts for its stall force that nearly exhausts[9] the fuel energy: $F_s \approx \Delta\mu/d \approx \Delta G_{min}(D(F=0))/d$.

The price for pure direction as a primary channel of energy consumption for nanomotors appears surprising at first glance, but is readily understood since $\Delta G_{min}(D)$ requires a few tens of $k_B T$, regardless of a motor's size, for a $D$ that is not apparently defective, but the energy consumption (i.e. work plus dissipation) processed by the motor per step quickly drops to that level with the motor's size shrinking to submicrometers. Hence the energy cost for pure direction is practically unique to nanoscale objects/machines, and becomes negligible for microscale or larger ones.

**IV. CONCLUSIONS**

In summary, quantifying pure direction with fidelity or directionality leads to a universal equality that captures the least possible energy price required by the 2nd law for sustaining the direction in an isothermal environment despite no work is done. Any lower price would violate the 2nd law in a thought experiment; real experimental proof comes from nanomotor measurements by a model-free comparison. A key to make high-performance nanomotors is to attain a high directional fidelity by the 2$^{nd}$-law decreed least energy price, as proven by biological nanomotor F1-ATPase that is known[9,14] to have ~ 100% efficiency for chemical energy-to-work conversion (i.e. $\Delta\mu \approx F_s d$). Indeed, a high directional fidelity per energy consumption amounts to an effective suppression of nonproductive futile steps and counter-productive backward steps of a motor, improving not only its energy utilization[5,6,18] but also other[2,3] performance like speed and force generation. The fidelity-energy equality indicates a threshold of energy input for fidelity improvement, and provides a general, quantitative guideline for design of nanoscale motors and devices.

## ACKNOWLEDGMENTS

We thank colleagues Jiangbin Gong, Jiansheng Wang for useful discussions. This work is partially supported by FRC grants under R-144-000-244-133 and R-144-000-259-112 (both to ZSW).

## REFERENCES

FIGURES

# Figure 1

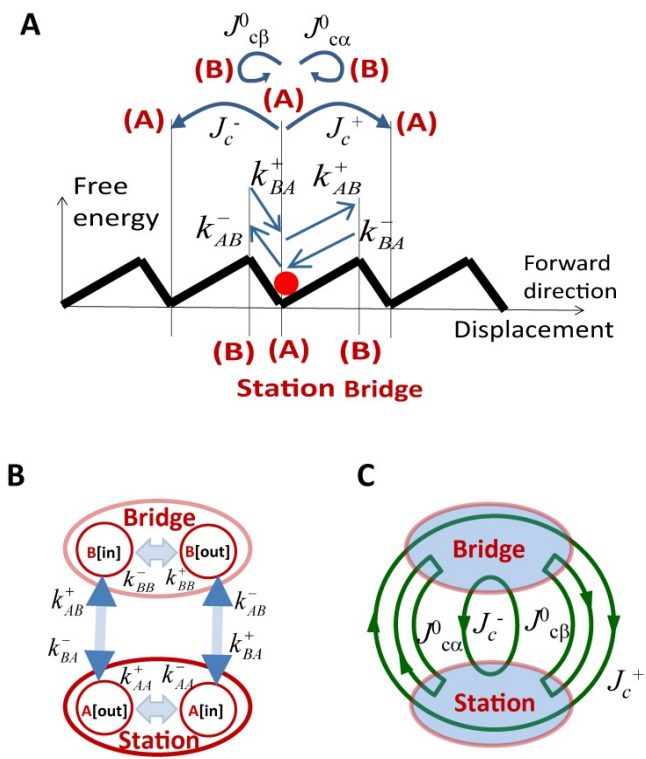

**Figure 1. Stochastic kinematics (A), transition representation (B), and elemental cycles (C) for an arbitrarily driven directional motion of a microscopic object in an isothermal environment.** A saw-toothed periodic potential is shown in **A** as an example for the environment-object interaction along the motional path of the object (sphere). In a displacement-free energy plot the path is characterized by an array of stations and bridges (**A**); and the motion is sufficiently described by transitions between a pair of adjacent station and bridge (**A, B**). The transitions form four elemental cycles (**A, C**) that quantify the directionality of a sustainable motion. $k$ is the transition rates and $J$ the cycle fluxes (superscripts $+$, $-$, $0$ mark transitions/cycles resulting in forward, backward and null displacements according to the motional direction).

# Figure 2

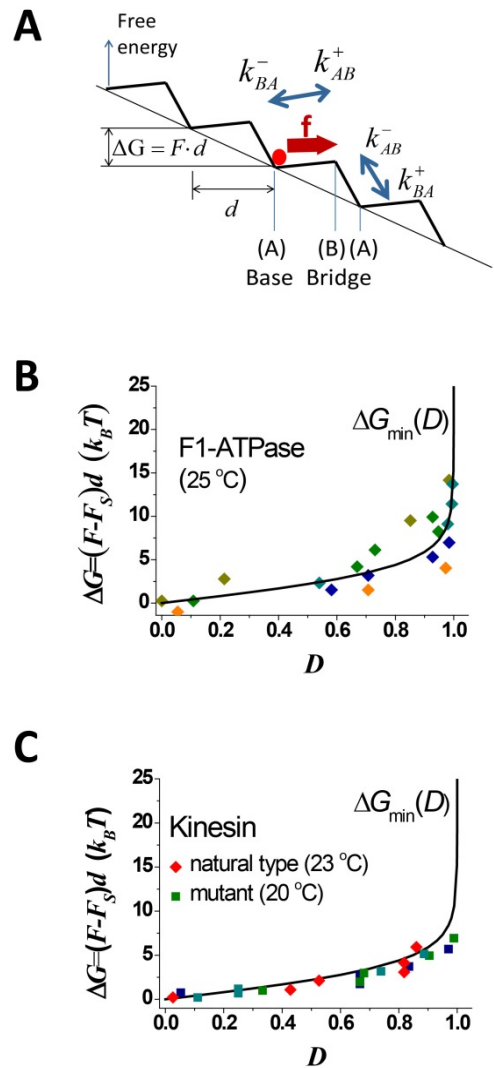

**Figure 2. The energy-direction equality versus experiments of force-induced directional motion. A**. Schematic illustration of a microscopic object under a constant pulling force ($F$) or equivalently an external field of constant slope in an isothermal environment represented by a saw-toothed potential of period $d$. The external field is superimposed to the environmental potential. The energy consumption is $\Delta G = F \times d$ per step of motion over distance $d$. **B, C**. Force-induced motion of biological nanomotors F1-ATPase (a rotational motor) and kinesin (a translational motor). A finite $D$ is produced by a pulling force or torque ($F$) that opposes the natural direction of the motors and overcomes the maximal resisting

force ($F_s$) that either motor produces by itself. The energy consumption driving such a reversed motion is $\Delta G = (F - F_s) \times d$ with the step size $d = 120°$ for F1-ATPase[14] and $d = 8.2$ nm for kinesin[15]. The equality (curve) is confronted with the $D$ versus $\Delta G$ data collected at different $F$ values for F1-ATPase (diamonds in **B**, from ref.[9], $F_s = 31.2$ pN nm/rad, different colours for data collected from different individual motor molecules), for natural kinesin (diamonds in **C**, from ref.[8], $F_s = 7.2$ pN for a saturating fuel concentration) and for mutant kinesin (squares, from ref.[10], $F_s = 4$ pN (blue), 3.2 pN (green), 2.4 pN (cyan) for three different fuel concentrations).

**Figure 3**

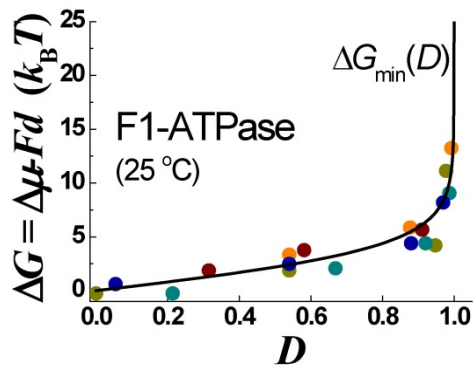

**Figure 3. The equality versus natural motion of biological nanomotor F1-ATPase.** This rotary motor produces a finite $D$ along its natural direction when it overcomes a small opposing torque $F$. The energy consumption driving the forward motion is $\Delta G = \Delta\mu - F \times d$ with $\Delta\mu$ as the chemical energy from the fuel consumption and $d$ the step size. The equality (curve) is confronted with the $D$ versus $\Delta G$ data collected at different $F$ values for F1-ATPase (filled circles, from ref.[9], colours mark data collected from different individual motor molecules; $\Delta\mu = 65.2$ pN nm and $d = 120°$).

1   E. R. Kay, D. Leigh, and F. Zerbetto, Angew. Chem. Int. Ed. **46**, 72 (2007);   W. B. Sherman and N. C. Seeman, Nano Lett. **4**, 1203 (2004);   J. S. Shin and N. A. Pierce, J. Am. Chem. Soc. **126**, 10834 (2004);   P. Yin, H. Yan, X. G. Daniell, A. J. Tuerberfield, and J. H. Reif, Angew. Chem. Int. Ed. **43**, 4906 (2004);   J. Bath, S. J. Green, and A. J. Turberfield, Angew. Chem. Int. Ed. **44**, 4358 (2005);   Y. Tian, Y. He, Y. Chen, P. Yin, and C. Mao, Angew. Chem. Int. Ed. **44**, 4355 (2005);   S. J. Green, J. Bath, and A. J. Turberfield, Phys. Rev. Lett. **101**, 238101(1 (2008);   T. Omabegho, R. Sha, and N. C. Seeman, Science **324**, 67 (2009);   M. von Delius, E. M. Geertsema, and D. A. Leigh, Nat. Chem. **2**, 96 (2009);   M. You, Y. Chen, X. Zhang, H. Liu, R. Wang, K. Wang, K. R. Williams, and W. Tan, Angew. Chem. Int. Ed. **51**, 2457 (2012);   J. Cheng, S. Sreelatha, R. Z. Hou, A. Efremov, R. C. Liu, J. R. van der Maarel, and Z. S. Wang, Phys. Rev. Lett. **109**, 238104 (2012).

2   A. Efremov and Z. S. Wang, Phys. Chem. Chem. Phys. **13**, 5159 (2011).

3   A. Efremov and Z. S. Wang, Phys. Chem. Chem. Phys. **13**, 6223 (2011).

4   R. D. Astumian, Science **276**, 917 (1997);   R. F. Fox, Phys. Rev. **E 57**, 2177 (1998);   M. E. Fisher and A. B. Kolomeisky, Proc. Natl. Acad. Sci. USA **96** (12), 6597 (1999);   R. Lipowsky, Phys. Rev. Lett. **85**, 4401 (2000);   P. Reimann, Physics Reports **361**, 57 (2002);   K. I. Okazaki, N. Koga, S. Takada, J. N. Onuchic, and P. G. Wolynes, Proc. Natl. Acad. Sci. USA **103**, 11844 (2006);   Z. S. Wang, Proc. Natl. Acad. Sci. USA **104**, 17921 (2007);   Z. S. Wang, M. Feng, W. W. Zheng, and D. G. Fan, Biophys. J. **93**, 3363 (2007);   Y. Xu and Z. S. Wang, J Chem Phys **131**, 245104(9) (2009);   C. Hyeon and J. N. Onuchic, Proc. Natl. Acad. Sci. USA **104**, 2175 (2007);   S. Rahav, J. Horowitz, and C. Jarzynski, Phys. Rev. Lett.

**101**, 140602(1 (2008)); V. Y. Chernyak and N. A. Sinitsyn, Phys. Rev. Lett. **101**, 160601(1 (2008)).

5   A. Parmeggiani, F. Julicher, A. Ajdari, and J. Prost, Phys. Rev. **E 60**, 2127 (1999).

6   G. Oster and H. Wang, Biochim. Biophys. Acta **1458**, 482 (2000).

7   R. D. Astumian, Proc. Natl. Acad. Sci. USA **102**, 1843 (2005).

8   N. J. Carter and R. A. Cross, Nature **435** (7040), 308 (2005).

9   S. Toyabe, T. M. Watanabe, T. Okamoto, S. Kudo, and E. Muneyuki, Proc. Natl. Acad. Sci. USA **108**, 17951 (2011).

10  B. E. Clancy, W. M. Behnke-Parks, J. O. L. Andreasson, S. S. Rosenfeld, and S. M. Block, Nat. Struct. Mol. Biol. **18**, 1020 (2011).

11  J. Schnakenberg, Rev. Mod. Phys **48**, 571 (1976); U. Seifert, Phys. Rev. Lett. **95**, 040602 (2005).

12  T. L. Hill, *Free Energy Transduction and Biochemical Cycle Kinetics* ((Springer, New York), 1989).

13  V. Bormuth, V. Varga, J. Howard, and E. Schaeffer, Science **325**, 870 (2009).

14  R. Yasuda, H. Noji, K. Kinosita, and M. Yoshida, Cell **93**, 1117 (1998).

15  M. Nishiyama, H. Higuchi, and T. Yanagida, Nature Cell Biology **4** (10), 790 (2002).

16  Y. Taniguchi, M. Nishiyama, Y. Ishii, and T. Yanagida, Nature Chem. Biol. **1** (6), 342 (2005).

17  A. B. Kolomeisky, E. B. Stukalin, and A. A. Popov, Physical Review E **71** (3), 031902 (2005); Q. Shao and Y. Q. Gao, Proc. Natl. Acad. Sci. USA **103** (21), 8072 (2006); S. Liepelt and R. Lipowsky, Phys. Rev. Lett. **98**, 258102 (2007); M. Bier, Eur. Phys. J. B **65**, 415 (2008); H. Y. Wang and C. J. He, Sci. China **54**, 2230 (2011).

[18] I. Derenyi, O. Bieri, and R. D. Astumian, Phys. Rev. Lett. **83**, 903 (1999); R. D. Astumian, Proc. Natl. Acad. Sci. USA **104**, 19715 (2007); U. Seifert, Phys. Rev. Lett. **106**, 020601 (2011); C. Van den Broeck, N. Kumar, and K. Lindenberg, Phys. Rev. Lett. **108**, 210602 (2012).